\documentclass[conference]{IEEEtran}
\usepackage{cite}
\usepackage{amsmath,amssymb,amsfonts}
\usepackage{algorithmic}
\usepackage{graphicx}
\usepackage{booktabs}
\usepackage{textcomp}
\usepackage{verbatim}
\usepackage{wasysym}

\usepackage[table]{xcolor}
\def\BibTeX{{\rm B\kern-.05em{\sc i\kern-.025em b}\kern-.08em
    T\kern-.1667em\lower.7ex\hbox{E}\kern-.125emX}}
\usepackage{tikz}
\usepackage{pgf-pie}
\usepackage{pgfplots}
\pgfplotsset{compat=1.17}
\usepackage{bbding}
\usepackage{enumitem}
\usepackage{colortbl}
\usepackage{multirow}
\usepackage[breaklinks=true]{hyperref}
\usepackage{comment}
\usepackage{dblfloatfix}

\newcommand\mytab[1]{\begin{tabular}[t]{c}#1\end{tabular}}
\newcommand\rotbox[1]{\rotatebox[origin=c]{90}{\mytab{#1}}}

\newcommand{\intervieweeone}{$i_1$}
\newcommand{\intervieweetwo}{$i_2$}
\newcommand{\intervieweethree}{$i_3$}
\newcommand{\intervieweefour}{$i_4$}
\newcommand{\intervieweefive}{$i_5$}
\newcommand{\intervieweesix}{$i_6$}
\newcommand{\intervieweeseven}{$i_7$}
\newcommand{\intervieweeeight}{$i_8$}
\newcommand{\intervieweenine}{$i_9$}

\newcommand{\apply}{$\CIRCLE$}
\newcommand{\applynoname}{$\CIRCLE^{\times}$}

\newcommand{\applydiff}{$\LEFTcircle$}
\newcommand{\noapply}{$\Circle$}
\newcommand{\noapplynoname}{$\Circle^{\times}$}

\newcommand{\mentioned}{\PencilRightDown}
\newcommand{\notmentioned}{\EightStarTaper}

\def\signed #1{{\leavevmode\unskip\nobreak\hfil\penalty50\hskip2em
  \hbox{}\nobreak\hfil--- #1%
  \parfillskip=0pt \finalhyphendemerits=0 \endgraf}}
\newsavebox\mybox

\makeatletter
\newcommand{\linebreakand}{%
  \end{@IEEEauthorhalign}
  \hfill\mbox{}\par
  \mbox{}\hfill\begin{@IEEEauthorhalign}
}
\makeatother

\begin{document}

\title{Designing Microservice Systems Using Patterns:\\ \huge{An Empirical Study on Quality Trade-Offs}
}

 \author{
     \IEEEauthorblockN{Guilherme Vale}
     \IEEEauthorblockA{
         \textit{\ \ \ \ \ Faculty of Engineering\ \ \ \ \ } \\
         \textit{University of Porto}\\
         Porto, Portugal \\
         up201709049@edu.fe.up.pt
     }
     \and
     \IEEEauthorblockN{Filipe Figueiredo Correia}
     \IEEEauthorblockA{
         \textit{\ \ \ \ \ Faculty of Engineering\ \ \ \ \ } \\
         \textit{University of Porto}\\
         \textit{INESCTEC}\\
         Porto, Portugal \\
         filipe.correia@fe.up.pt
     }
     \and
     \IEEEauthorblockN{Eduardo Martins Guerra}
     \IEEEauthorblockA{
         \textit{Computer Science Faculty} \\
         \textit{Free University of Bozen-Bolzano}\\
         Bolzano, Alto Adige, Italy \\
         eduardo.guerra@unibz.it
     }
     \linebreakand
     \IEEEauthorblockN{Thatiane de Oliveira Rosa}
     \IEEEauthorblockA{
         \textit{Institute of Mathematics and Statistics} \\
         \textit{University of São Paulo}\\
         São Paulo, SP, Brazil \\
         thatiane@ime.usp.br
     }
     \and
     \IEEEauthorblockN{Jonas Fritzsch}
     \IEEEauthorblockA{
         \textit{\ \ \ Institute of Software Engineering\ \ \ } \\
         \textit{University of Stuttgart}\\
         Stuttgart, Germany \\
         jonas.fritzsch@iste.uni-stuttgart.de
     }
     \and
     \IEEEauthorblockN{Justus Bogner}
     \IEEEauthorblockA{
         \textit{\ \ \ Institute of Software Engineering\ \ \ } \\
         \textit{University of Stuttgart}\\
         Stuttgart, Germany \\
         justus.bogner@iste.uni-stuttgart.de
     }
 }

\maketitle

\begin{abstract}
The promise of increased agility, autonomy, scalability, and reusability has made the microservices architecture a \textit{de facto} standard for the development of large-scale and cloud-native commercial applications. 
Software patterns are an important design tool, and often they are selected and combined with the goal of obtaining a set of desired quality attributes. However, from a research standpoint, many patterns have not been widely validated against industry practice, making them not much more than interesting theories. 
To address this, we investigated how practitioners perceive the impact of 14 patterns on 7 quality attributes. Hence, we conducted 9 semi-structured interviews to collect industry expertise regarding (1) knowledge and adoption of software patterns, (2) the perceived architectural trade-offs of patterns, and (3) metrics professionals use to measure quality attributes.
We found that many of the trade-offs reported in our study matched the documentation of each respective pattern, and identified several gains and pains which have not yet been reported, leading to novel insight about microservice patterns.
\end{abstract}


\section{Introduction}

Designing successful microservices-based applications requires mastering a set of new architectural insights, principles, and practices. 
These aspects are, however, still far from being fully understood.
One such aspect is the use of patterns; as professionals encounter and overcome challenges that arise from this architectural approach, shared solutions to specific problems begin to emerge.
The documentation of these patterns gives software architects an array of techniques to common hurdles, thus reducing the technical risk to their projects by not having to employ new and untested designs.

On the other hand, patterns are not \textit{inherently} tried-and-true~\cite{kohls2009true}. Although they are usually documented when they have been observed in practice at least three times, such a \textit{convenience sample}~\cite{baltes2020sampling} often does not provide enough evidence of their validity. Therefore, there is value in subjecting patterns to further evaluation, like with any other type of theory~\cite{riehle2021pattern}. 

The promises of microservices---\textit{e.g.}, of increased agility, scalability, maintainability, performance---can only be leveraged with a careful understanding of its underlying principles.
Amid the hype created in recent years, as many big industry players pushed this architecture~\cite{carvalho_analysis_2019}, the approaches that work best for adopting it are still an object of study.
For this reason, it is not uncommon to find contexts wherein microservices would be a good fit, but where teams fail to implement them successfully~\cite{jamshidi}.
Discovering how industry experts currently perceive the influence of design patterns on quality attributes (QAs) will thus help to judge their benefits and relevance, which forms the motivation for this work.

\section{Research approach} \label{sec:rq}
\label{sec:approach}

We seek a more in-depth understanding of microservice design patterns by way of an empirical study of their trade-offs. 
Specifically, we want to obtain insights on the relevance of design patterns in industry, how practitioners perceive their influence on software qualities as a consequence of their usage, and what metrics practitioners use, if any, to determine these derived effects, reflected as software qualities. 
Based on these objectives, we defined the following research questions (RQs).

\begin{itemize}[leftmargin=*]
    \item \textbf{RQ1:} \textbf{\textit{What is the rationale for the adoption of patterns in microservices systems?}} 
    --- 
    Considerations: how relevant are microservice design patterns in industry? What is their utility? Are developers aware of the problems that benefit from the recommended usage of these patterns?
    
    \item \textbf{RQ2:} \textbf{\textit{How are QAs influenced as a result of applying microservice patterns?}}
    --- 
    Considerations: what are the trade-offs, in terms of advantages and disadvantages, of current microservice patterns as perceived by professionals?
    
    \item \textbf{RQ3:} \textbf{\textit{How are QAs measured in microservices?}} 
    --- 
    Considerations: do practitioners evaluate their non-functional requirements? If so, how? With which techniques and resources? If not, why?
\end{itemize}

To answer these RQs, we opted for a semi-structured interview study with microservices experts, as face-to-face interviews can help us establish a more effective communication channel with participants.

\subsection{Research scope} \label{sec:scope}

Several patterns have been described in the scope of the microservices architecture and cloud-native development. We sought to select a set of patterns as a basis for our study that would be broad, consistent, detailed, and regarded as a reference. Some works, such as Richardson's~\cite{richardson_microservices_2019} do not describe the patterns themselves, or when they do, the descriptions vary widely in depth, often not identifying the consequences inherent to applying the patterns\footnote{In addition to the book, which does not describe the patterns themselves, Richardson makes a few patterns available at \url{https://microservices.io}.}. Other pattern languages did not seem to be sufficiently broad~\cite{balalaie2018microservices,marquez2018pattern,sousa2018autorecovery,sousa2018extmonitoring,sousa2017messaging,sousa2015contanerization} or were in very early and active development~\cite{cloudpatterns2021brown}.

Considering our focus on industry-relevance, we selected a set of patterns from the Azure Architecture Center (AAC), a repository of software architecture documentation hosted by Microsoft.
The AAC has an extensive language of patterns termed the \textit{Cloud Design Patterns}~\cite{cdp_aac}, divided into 7 categories, with most patterns belonging to more than one category. In particular, we selected the patterns of the \textit{design and implementation} category, as they address the most general aspects of designing a microservices architecture. 

The categorization and taxonomy of architectural patterns for microservices varies significantly in literature: Márquez and Astudillo \cite{marquez2018actual} assigned 17 architectural patterns used in open-source projects to 11 groups while Valdivia et al. \cite{valdivia_patterns_2020} identified 54 patterns from white and grey literature and divided them into 6 groups. The taxonomy suggested by Osses et al. \cite{Osses2018a} who reported 124 architectural patterns in the academic and industrial field has the greatest overlap with Microsoft's AAC. 8 of the 14 patterns in the AAC \textit{design and implementation} category were also tagged as design patterns by Osses et al. while another one occurs as a migration pattern. 

The patterns in the AAC are available in a public GitHub repository\footnote{Available at \url{https://github.com/MicrosoftDocs/architecture-center}} that is actively maintained and has frequent community contributions. We have analyzed 14 specific patterns (\textit{cf}. Section~\ref{sec:findings}) in the version corresponding to the commit with hash \texttt{eff408a}, from June 11th 2021. 

As for QAs, we turned to the work by Li \textit{et al.}~\cite{li_understanding_2021}, which dealt with questions of assessing QAs in microservice systems, as it matched our desired level of architectural perspective. 
The authors produce a list of six QAs most relevant to the microservices architecture.
For our part, we extend this set of QAs to include a seventh element, \textit{maintainability}. 
The list that follows contains the full set of QAs of our study, a brief exploration of each, and a description of their relevance to the microservices architecture.

\begin{enumerate}[leftmargin=*]
    \item[] \textbf{Scalability}. A measure of a system's ability to handle a varying number of requests~\cite{bass2003software}. 
    A microservices architecture is often adopted to maximize aspects of scalability, given its proximity to cloud-native technologies and ability to simplify both vertical and horizontal scaling techniques.
    
    \item[] \textbf{Performance}. A measure of a system's ability to meet time requirements when responding to inputs, requests, or events~\cite{bass2003software}.
    As a distributed architecture, the latency in network requests involved in inter-service communication is an obvious source of performance hindrance.
    Established lightweight and REST-based communication mechanisms are commonly used to mitigate this threat.
    
    \item[] \textbf{Availability}. A measure of a system's ability to repair faults within a defined period so as not to incur significant performance penalties~\cite{bass2003software}.
    As such, we can think of \textit{availability} as interchangeable with \textit{reliability}~\cite{li_understanding_2021}.
    Since services typically have at least one dependency, downtime in one faulty service can easily have implications that extend to otherwise well-behaved services. 
    Several patterns for addressing availability are known, most notably circuit breakers, fault monitors, and service registries.
    
    \item[] \textbf{Monitorability}. A measure of a system's ability to support the operations staff in monitoring the system at runtime, \textit{i.e.}, while it is executing~\cite{bass2003software}. 
    Any distributed system, conceived as a microservice system or otherwise, exhibits a high level of dynamic structure and behavior, representing an added challenge for monitoring~\cite{li_understanding_2021}. 
    This QA is often also referred to as \textit{observability}. 
    We use the two terms interchangeably throughout this work.
    
    \item[] \textbf{Security}. A measure of a system's ability to protect data from unauthorized access while enabling access to authorized users and systems~\cite{bass2003software}. 
    The distribution of application logic into different processes results in complex network interaction models between services~\cite{li_understanding_2021}. As a result, this added complexity can be more easily exploited by attackers if \textit{security} is neglected. 
    
    \item[] \textbf{Testability}. A measure of a system's ability to demonstrate its faults through testing---typically execution-based testing~\cite{bass2003software}.
    Due to the complex relationship between services, \textit{testability} is an important property that helps to detect and minimize impacts on other qualities like \textit{performance}, \textit{security} or \textit{availability}~\cite{li_understanding_2021}.
    
    \item[] \textbf{Maintainability}.
    It concerns the balancing act between the time and resources required to apply changes and the need to introduce new business logic or improve system functionality~\cite{bass2003software}. 
    Poor design can easily inflate the cost of change. 
    The coupling of different services may imply that changes to a service cannot be realized without orchestrating similar changes in services depending on it, which is a sign of poor maintainability.
\end{enumerate}

\subsection{Interview design} \label{sec:structure}

\begin{table*}[b]
\begin{center}
    \caption[Participant demographics]{Participant demographics.}
    \label{tab:participants}
    \begin{tabular}{cllll}
    \toprule
    Interviewee         & Role                  & Years of experience   & Years with microservices  & Industry      \\ 
    \midrule
    \intervieweeone       & Architect             & 10 to 14              & 5 or more                 & Artificial Intelligence            \\
    \intervieweetwo      & Chief Technology Officer & 10 to 14              & 2 to 4                    & Retail        \\
    \intervieweethree      & Operations            & 5 to 9                & 2 to 4                    & Artificial Intelligence            \\
    \intervieweefour       & QA                    & 5 to 9                & 2 to 4                    & Consultancy   \\
    \intervieweefive      & Scrum Master          & 5 to 9                & 5 or more                 & Automotive    \\
    \intervieweesix          & Chief Technology Officer & 15 to 19              & 2 to 4                    & Logistics     \\
    \intervieweeseven    & Developer             & 5 to 9                & 2 to 4                    & Gaming        \\
    \intervieweeeight      & Head of Engineering   & 10 to 14              & 2 to 4                    & Finance       \\
    \intervieweenine          & Delivery Manager      & 10 to 14              & 5 or more                 & Finance       \\
    \bottomrule
    \end{tabular}
\end{center}
\end{table*}

Semi-structured interviews help to explore both planned-ahead and unforeseen topics.
We used the advice of Hove and Anda~\cite{hove2005} and of Seaman~\cite{Seaman2008} for planning and conducting the interviews. As well, we considered the guidelines by Ralph \textit{et al.}~\cite{ralph2021empirical}, which provide checklists that help to ensure scientific rigor for empirical studies in software engineering.

All 9 interviews followed the below outlined three-part structure, that was initiated by general and preparatory remarks, \textit{i.e.}, greetings, personal presentation, study presentation and obtaining authorization for recording:

\begin{enumerate}[leftmargin=*]
    \item[] \textbf{Part 1:} \textbf{Personal experience with microservices.} 
    We ask the participant to elaborate freely on their professional experience with microservices. 
    We ask them to consider details like the number of services that were in development and operation in these projects, the number of lines of code that composed the system, the service communication mechanisms used, and the deployment strategies used. 
    
    The goal of this part is to understand the microservice systems the participant has worked with, helping to contextualize their answers in the rest of the interview.
    
    \item[] \textbf{Part 2:} \textbf{Design pattern trade-offs.} 
    We present the participant with the design patterns we selected in Section~\ref{sec:scope}, using a helper slideshow document. Each pattern was described by a different slide, with its name, brief description, and a figure. 
    The figures are taken directly from the respective documentation of the patterns in the AAC. 
    In any case, for each pattern, we show its corresponding slide, briefly describe it to the participant and then immediately ask if the participant has ever applied this pattern. 
    If the answer is no, we skip to the next pattern. If the answer is yes, we ask open-ended questions about the decisions that motivated the adoption of the pattern and what they felt was improved and worsened, in a global sense: not necessarily what was impacted on a localized level (\textit{e.g.}, per service) but on a generalized level. 
    
    The goal of this part is twofold: to understand the rates of knowledge and adoption of the patterns, and to understand the trade-offs practitioners perceive as inherent to each. 
    
    \item[] \textbf{Part 3:} \textbf{Microservice quality metrics.} 
    We present the participant with the QAs we selected in Section~\ref{sec:scope}. 
    We then ask the participant to go through the set of QAs, in any order they choose, and to tell us, for each attribute, how much of a concern it is for their company, how they measure it, and how often they measure it. 
    
    The goal of this part is to determine the perceived importance of software qualities, plus the metrics used to evaluate them. 
\end{enumerate}

\subsection{Sampling strategy} \label{sec:sampling} \label{sec:interviews}

We chose to use \textit{purposive sampling}~\cite{baltes2020sampling} and reached out to colleagues or associates in the industry who either served as participants or forwarded us to potential participants. 
We also attempted \textit{heterogeneity sampling} by contacting candidates from a breadth of different industry domains and performing different roles. 
All in all, we have interviewed, over a period of two and a half months, 9 professionals from 9 different companies; all with development offices in Portugal.
An overview of the participants is shown in Table~\ref{tab:participants}.

The interviewees work in widely different industries, with only 2 industry domains appearing more than once---\textit{artificial intelligence} and \textit{finance}.
All participants have at least 5 years of experience with software development.
A third of them (3/9) reported 5 or more years of experience with microservices, the remaining two thirds (6/9) have between 2 and 4 years of experience with microservices.
Some participants are in roles that play a strong part in defining the system architecture (\textit{i.e.}, CTO, architect, head of engineering), others require, at the very least, a sound understanding of the architecture (\textit{i.e.}, operations, developer, delivery manager). The remaining two had professional profiles and \textit{de facto} roles that went beyond their official role (\textit{i.e.}, \intervieweefive\ often is involved in technical leadership, and \intervieweefour\ is concerned with internal quality aspects, which extend to the architecture of the system).

The transcription of the interviews was done manually using the \textit{oTranscribe} tool. A total of $47\,322$ words were transcribed from 537 minutes and 40 seconds of recorded video.

\subsection{Data analysis} \label{sec:analysis}

The data analysis in this study was performed with a combination of manual and computer-assisted techniques: we used the qualitative data analysis tool NVivo for coding the large amount of text data we obtained, and we manually identified themes and other findings into a structured form. In an ongoing process parallel to data collection, we performed initial open theory-driven coding, a process of breaking data apart and defining boundaries to stand for blocks of raw data~\cite{Stol2016}.
By envisioning a set of expected topics that we had foreseen \textit{a priori} from our RQs, we developed the top-level codes \textit{Preliminary}, \textit{Design Patterns}, \textit{Measuring QAs}, and \textit{Miscellaneous} and a code for every pattern and QA.
These form the structural areas of our analysis. 
Through later axial coding, a higher level of coding~\cite{decuir2011developing}, we establish links between the codes through deductive reasoning: each pattern is a subcategory of the \textit{Design Patterns} code, similarly for each QA being a subcategory of \textit{Measuring QAs}.

\begin{figure}[htbp]
\centering
\includegraphics[width=\linewidth]{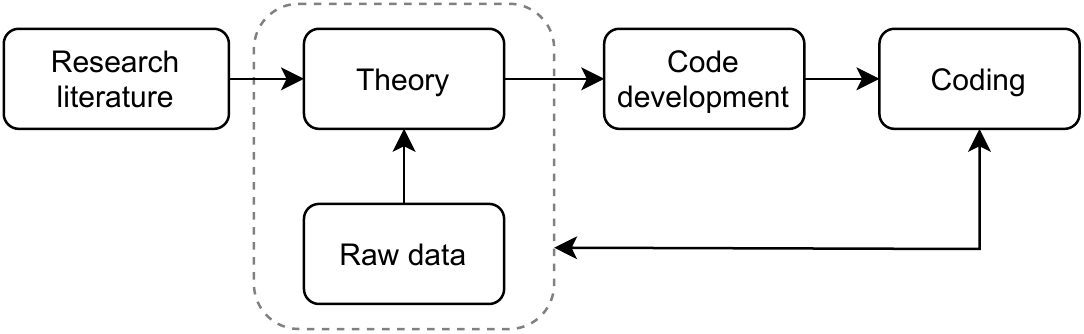}
\caption[Coding process]{Coding process.}
\label{fig:coding-process}
\end{figure}

With our raw data extracted, we assigned these codes to reduce and simplify our data. 
When suitable, we expanded the data (forming new connections between concepts) or reconceptualized it (rethinking our previous theoretical assumptions). 
At the same time, we examined how the theory guiding our research was supported or contradicted by the data, as well as the impact of the data on the current research literature.
An overview of this process is shown in Figure~\ref{fig:coding-process}. We make available the materials used by the researchers to conduct the study, as well as aggregated and anonymized data produced in this context~\cite{guilherme_zenodo_2021}.

\begin{table*}[t]
\caption[Reported patterns adopted by interviewees]
{
Reported patterns adopted by interviewees, ordered by frequency of adoption (\#).
}
\centering
\label{tab:pattern-adoption}

\begin{minipage}[t]{0.74\linewidth}
\vspace*{-\dimexpr\baselineskip+\heavyrulewidth+\abovetopsep\relax} % magic. this is used to vertically align the table and the legend to the right
\quad\begin{tabular}{ll@{\hskip 0.04in}l@{\hskip 0.04in}l@{\hskip 0.04in}l@{\hskip 0.04in}l@{\hskip 0.04in}l@{\hskip 0.04in}l@{\hskip 0.04in}l@{\hskip 0.04in}l@{\hskip 0.15in}c}
\toprule
Pattern                         & \intervieweeone & \intervieweetwo & \intervieweethree & \intervieweefour & \intervieweefive & \intervieweesix & \intervieweeseven & \intervieweeeight & \intervieweenine & \# \\ \midrule

\textsc{Gateway Routing}                & \apply        & \apply        & \apply            & \apply        & \apply            & \apply  & \apply   & \apply   & \apply &9 \\
\textsc{Strangler}                      & \apply        & \apply        & \apply            & \apply        & \apply            & \apply  & \applynoname & \applynoname & \applynoname & 9\\
\textsc{External Configuration Store}   & \apply        & \apply        & \apply            & \noapply      & \apply            & \apply  & \applynoname   & \apply   & \apply & 8\\
\textsc{Static Content Hosting}         & \apply        & \apply        & \apply            & \noapply      & \apply            & \apply  & \apply   & \apply   & \apply & 8\\ 
\textsc{Pipes and Filters}              & \apply        & \apply        & \apply            & \applynoname        & \apply            & \apply  & \apply   & \noapply  & \apply & 8\\
\textsc{Command Query Responsibility Segregation}                           & \apply        & \apply        & \noapply          & \noapply      & \applydiff    & \apply  & \applynoname  & \applynoname   & \apply & 6\\
\textsc{Gateway Offloading}             & \apply        & \apply        & \apply            & \noapply      & \apply            & \apply  & \noapply  & \apply   & \noapply & 6 \\
\textsc{Backends for Frontends}         & \apply    & \apply        & \noapply          & \noapply      & \apply            & \apply  & \noapply  & \noapply  & \apply & 5\\
\textsc{Compute Resource Consolidation} & \applynoname  & \applynoname  & \noapply          & \noapply      & \applynoname      & \apply  & \noapplynoname   & \noapply  & \applynoname & 5 \\
\textsc{Sidecar}                        & \apply        & \apply        & \noapply          & \apply        & \apply            & \apply & \noapply  & \noapply  & \noapply & 5 \\
\textsc{Anti-Corruption Layer}          & \noapply        & \apply        & \noapplynoname    & \apply        & \noapply          & \noapplynoname & \noapply  & \applynoname  & \apply & 4 \\
\textsc{Ambassador}                     & \apply        & \apply        & \applydiff          & \noapply      & \apply            & \noapply & \noapply   & \applydiff   & \noapply & 3 \\
\textsc{Gateway Aggregation}            & \apply        & \apply        & \noapply          & \noapply      & \apply            & \noapply & \noapply  & \noapply  & \noapply & 3 \\
\textsc{Leader Election}                & \noapply        & \apply        & \noapply            & \noapply      & \apply            & \noapply  & \apply   & \noapplynoname  & \noapply & 3 \\
\midrule
Number of patterns adopted by participant & \hspace{-0.01in}12 & \hspace{-0.01in}14 & \hspace{0.02in}6 & \hspace{0.02in}5 & \hspace{-0.01in}12 & \hspace{-0.01in}10 & \hspace{0.02in}7 & \hspace{0.02in}7 & \hspace{0.02in}9 & \\

\bottomrule
\end{tabular}
\end{minipage}\begin{minipage}[t]{0.26\linewidth}
\vspace{1em}
\begin{itemize}
\scriptsize
    \setlength\itemsep{0.4em}
    \item[\apply\ ~] The pattern was said to have been applied by the participant; 
    \item[\applynoname] The pattern was said to have been applied, but the participant was not familiar with the given pattern's name; 
    \item[\noapply\ ~] The pattern was said to not have been applied by the participant;
    \item[\noapplynoname] The pattern was said to have neither applied nor even known to the participant
    \item[\applydiff\ ~] The pattern was said to have been applied but the participant's description of their usage of the pattern does not match the intent originally documented.
\end{itemize}\hspace{0.5em}

\end{minipage}

\end{table*}

\section{Results} \label{sec:findings}

The following sections describe our results and explore their implications and limitations.
We map the first three sections to our RQs, in Sections~\ref{sec:rationale}, \ref{sec:tradeoff} and \ref{sec:measuring}, respectively.
In Section~\ref{sec:threats} we discuss the threats to the validity of our results.
We present these findings almost always in a table format, with a column ``\textit{\#}" representing the number of unique respondents that mentioned a particular theme, \textit{i.e.}, the number of occurrences of a given idea.

\subsection{Rationale for pattern adoption} \label{sec:rationale}

Table~\ref{tab:pattern-adoption} presents the reported adoption of patterns by our participants. The average number of patterns adopted by a participant is 9. 
Only 1 of the 9 participants (\intervieweetwo) reported having had experience with all 14 patterns.
Only 2 patterns were applied by all, \textsc{Strangler} and \textsc{Gateway Routing}, with the former receiving particular praise from most respondents.

Many participants recognized many of the patterns but had not implemented them themselves. 
The \textsc{Leader Election} pattern is probably the best to illustrate this point: all but one participant were aware that leader election is a strategy used in distributed systems, \textit{e.g.}, in database management systems and container orchestration mechanisms, but only three reported having actually implemented it in their products.

In some instances, participants responded with having adopted a pattern but, upon further discussion regarding their application, we concluded that it was not an application of the pattern at hand. 
Two participants erroneously said they applied the \textsc{Ambassador} pattern, confusing it with a gateway.

Additionally, one participant interpreted the \textsc{CQRS} pattern as referring simply to the usage of read-only database replicas to improve performance.
While using read replicas is one way to set up a read store with this pattern, a crucial component is the separation of a data model in two, one for updates (commands) and one for reads (queries). 
We assign these instances a \applydiff\ value.
These most likely reflect miscommunication or misunderstandings; instances of our inability to convey the essence of a pattern to respondents. 
As a result, in these cases, we ignore the participants' perceptions of trade-offs to not jeopardize or corrupt our findings in the next section.

\subsection{Pattern trade-offs} \label{sec:tradeoff}

In the following sections, we explore the findings related to the professional perception of the software qualities affected by each pattern.
We express the trade-offs of the patterns through the form of \textit{gains} and \textit{pains}. Table~\ref{tab:pattern-tradeoffs} summarizes the trade-offs as reported by participants for each pattern.
We attach one of two symbols to each concern raised by respondents. 
The \mentioned\ symbol denotes remarks which are already identified in the AAC documentation, and the \notmentioned\ symbol denotes novel remarks which are not present in the AAC. 

We note here that, when asked for advantages and disadvantages, participants rarely invoked QAs directly, referring instead to benefits and liabilities in terms of concerns raised at design, development or operation times.
As such, the findings mix QAs, when they were mentioned explicitly in responses, and the aforementioned procedures or properties that were relayed to us. 

\subsubsection{\textsc{Strangler}}

Applied by all nine participants. Over half of them mention \textbf{scalability} as a gain received from applying this pattern, with \textbf{maintainability} trailing close by. 
Additionally, the fact that the strangler façade allows developers to keep working on migrating to microservices without having to demand adjustments from clients---what we term, for our purposes, as \textbf{consumer transparency}---was a major advantage that practitioners mentioned.
As for downsides, the \textbf{costs to maintain} the arrangement this pattern demands and the careful \textbf{planning of transition periods} required to ensure successful migrations received most mentions. 
One participant reported difficulties \textbf{managing data consistency} between subsystems, and another reported \textbf{initial resistance by the development team} when presenting the idea for this pattern.
Finally, another participant noted \textbf{backwards compatibility} as a particular challenge introduced by this pattern. 

\subsubsection{\textsc{Anti-Corruption Layer}}

Applied by four participants. All of them note the same advantage with this pattern: the \textbf{decoupling between systems}.
On the other hand, the \textbf{complexity of the layer} was mentioned by three participants as a serious drawback to be considered. 
Additionally, a negative impact on \textbf{performance} was noted by one participant.

\subsubsection{\textsc{Sidecar}}

Applied by five participants. All of them noted \textbf{monitorability} as the major gain of this pattern.
\textbf{Maintainability} also followed in the set of benefits of sidecars. 
One participant additionally suggested \textbf{scalability} and \textbf{security} as further gains from this pattern.
Additionally, one participant mentioned \textbf{testability} as a facet enhanced by sidecars, with an interesting use case to illustrate.
Participants seemed happy with this pattern, with only one pointing out \textbf{operational complexity} as a pain brought about by its adoption.

\subsubsection{\textsc{Ambassador}}

Applied by three participants. \textbf{Availability} was praised by all three participants as a major gain from this pattern, as it enables easy implementation of resilience mechanisms.
Additionally, the proxy nature of the ambassador layer allows for greater \textbf{security}, as reported by two participants.
Similar to \textsc{Sidecar}, \textbf{monitorability}, \textbf{maintainability} and \textbf{scalability} were also cited as benefits of this pattern. 
All three participants reported implementing this pattern as a sidecar.
\textbf{Operational complexity} was mentioned by all three participants as the only downside to this pattern.

\subsubsection{\textsc{Compute Resource Consolidation}}

Applied by five participants. All of them noted \textbf{scalability} as the QA most helped by the application of this pattern, praising the autoscaling solutions which provide an easy interface for sophisticated scaling strategies, calculating the best consolidation of instances automatically. 
One participant noted that the scalability enabled by this pattern adds greatly to the serverless functions they use, as the serverless paradigm encourages processing on-demand rather than always running and waiting for requests.
Additionally, with the conjunction of sophisticated techniques like autoscaling, \textbf{cost savings} were noted by two participants as a welcome addition. 
No participant reported pains with this pattern. 
This suggests that participants consider this pattern to be a boon with minimal to no downsides.

\subsubsection{\textsc{Command Query Responsibility Segregation}}

Applied by six participants. They note gains in \textbf{scalability}, \textbf{performance} and \textbf{availability} as a result of this pattern.
As for drawbacks, the \textbf{complexity} introduced was mentioned by all but one participant.
Finally, \textbf{ensuring data consistency} was another problem felt by 2 participants.

\subsubsection{\textsc{External Configuration Store}}

Applied by eight participants. Our results suggest a consensus among our participants: the \textbf{decoupling between development and operational concerns} and \textbf{maintainability} are the primary gains of this pattern.
Additional benefits reportedly include \textbf{security}, \textbf{scalability} and \textbf{availability}.
On the other hand, the \textbf{complexity} of this setup should be taken into account, according to three participants.
Additionally, the external store represents a \textbf{single point of failure}, as mentioned by two participants.
Finally, one participant mentioned \textbf{difficulties introducing breaking changes} with this pattern present.

\subsubsection{\textsc{Gateway Routing}}

Applied by all nine participants. Participants see this pattern straightforwardly.
The fact that the gateway functions provides a \textbf{decoupling between the client and the services} through a single access layer was a gain reported by seven participants.
However, a gateway represents a \textbf{single point of failure}, a concern raised by four participants.

\subsubsection{\textsc{Gateway Aggregation}}

Applied by three participants. The three have noted that \textbf{performance} was the main reason behind choosing this pattern, as it allows for client applications to spend less time communicating with the API gateway.
However, none were particularly pleased with this solution. 
Similar to the previous pattern, the gateway remains a \textbf{single point of failure}. 
Furthermore, negative impacts in the forms of difficulties \textbf{scalability} of the gateway and possibilities of \textbf{coupling} between the gateway and the business microservices were noted by participants.

\subsubsection{\textsc{Gateway Offloading}}

Applied by six participants. Widely praised by all who adopted it, this pattern gives practitioners ways of tackling \textbf{security} foremost, followed by \textbf{maintainability} and \textbf{monitorability}.
One participant noted \textbf{performance} as an extra gain, which logically followed from the security benefits gained.
As for drawbacks, only one participant mentioned \textbf{difficulties defining which cross-cutting concerns to share} in the gateway.

\begin{table*}[t]
\scriptsize
\centering
\caption{Reported gains and pains of the patterns considered in the study.\\  The symbols show which items were already present in the AAC documentation (\mentioned) or were not present (\notmentioned).}
\label{tab:pattern-tradeoffs}
    \begin{tabular}[t]{
    >{\centering\arraybackslash}p{0.4cm}<{\hspace{-0.5em}}p{3cm}lp{3.1cm}l}
    & \textbf{Gains} & \# & \textbf{Pains} & \# \\
    
    \cmidrule(lr){2-3} \cmidrule(lr){4-5}
    \multirow{5}{*}{\rotbox{Strangler}}
    & \notmentioned\ Scalability        & 5 & \notmentioned\ Costly to maintain               & 2 \\
    & \notmentioned\ Maintainability    & 4 & \mentioned\ Planning transitions                & 2 \\
    & \mentioned\ Consumer transparency & 4 & \mentioned\ Managing data consistency           & 1 \\
    &                                   &   & \notmentioned\ Initial developers resistance    & 1 \\
    &                                   &   & \mentioned\ Backwards compatibility             & 1 \\[0.5em]

    \cmidrule(lr){2-3} \cmidrule(lr){4-5}
    \multirow{3}{*}[1.4em]{\rotbox{Anti-Corrup-\\ tion Layer}}
    & \mentioned\ Decoupling systems              & 4 & \mentioned\ Layer complexity & 3 \\
    &                                             & & \mentioned\ Performance    & 1 \\
    &                                             & & &  \\[0.5em]
    
    \cmidrule(lr){2-3} \cmidrule(lr){4-5}
    \multirow{5}{*}{\rotbox{Sidecar}}
    & \mentioned\ Monitorability    & 5  & \mentioned\ Operational complexity & 1  \\
    & \mentioned\ Maintainability   & 4  &                        &    \\
    & \notmentioned\ Scalability    & 1  &                        &    \\
    & \notmentioned\ Security       & 1  &                        &    \\
    & \notmentioned\ Testability    & 1  &                        &    \\[0.5em]
    
    \cmidrule(lr){2-3} \cmidrule(lr){4-5}
    \multirow{5}{*}{\rotbox{Ambassador}}
    & \mentioned\ Availability       & 3  & \mentioned\ Operational complexity & 3  \\
    & \mentioned\ Monitorability     & 2  &                        &    \\
    & \mentioned\ Security           & 2  &                        &    \\
    & \notmentioned\ Maintainability & 1  &                        &    \\
    & \notmentioned\ Scalability     & 1  &                        &    \\[0.5em]

    \cmidrule(lr){2-3} \cmidrule(lr){4-5}
    \multirow{5}{*}[1.2em]{\rotbox{Compute Resource\\ Consolidation}}
    & \mentioned\ Scalability     & 5  & \multicolumn{1}{c}{---}      &    \\
    & \mentioned\ Cost savings    & 2  &          &       \\
    & & & &\\
    & & & &\\
    & & & &\\
    & & & &\\[0.7em]

    \cmidrule(lr){2-3} \cmidrule(lr){4-5}
    \multirow{3}{*}{\rotbox{CQRS}}
    & \mentioned\ Scalability  & 3  & \mentioned\ Complexity       & 5  \\
    & \mentioned\ Performance  & 2  & \mentioned\ Ensuring data consistency & 2  \\
    & \notmentioned\ Availability & 2  &                           &    \\[0.5em]

    \cmidrule(lr){2-3} \cmidrule(lr){4-5}
    \multirow{5}{*}[1em]{\rotbox{External Confi-\\guration Store}}
    & \mentioned\ Decoupling      & 6  & \mentioned\ Complexity                     & 3  \\
    & \mentioned\ Maintainability & 6  & \mentioned\ Single point of failure                 & 2  \\
    & \notmentioned\ Security     & 2  & \notmentioned\ Making breaking changes & 1  \\
    & \notmentioned\ Scalability  & 1  &                                         &    \\
    & \notmentioned\ Availability & 1  &                                         &    \\[0.5em]
    
    \cmidrule(lr){2-3} \cmidrule(lr){4-5}
    \end{tabular}
\quad
    \begin{tabular}[t]{
    >{\centering\arraybackslash}p{0.4cm}<{\hspace{-0.5em}}p{3cm}lp{3.1cm}l}
     & \textbf{Gains} & \# & \textbf{Pains} & \# \\

    \cmidrule(lr){2-3} \cmidrule(lr){4-5}
    \multirow{2}{*}[1em]{\rotbox{Gateway\\ Routing}}
    & \mentioned\ Decoupling & 7  & \mentioned\ Single point of failure      & 4  \\
    & & & &\\
    & & & &\\[0.9em]
    
    \cmidrule(lr){2-3} \cmidrule(lr){4-5}
    \multirow{3}{*}[1em]{\rotbox{Gateway\\ Aggregation}}
    & \mentioned\ Performance & 3  & \mentioned\ Single point of failure      & 2  \\
    & &    & \mentioned\ Scalability                  & 2  \\
    & &    & \mentioned\ Coupling              & 1  \\
    & & & &\\[0.9em]

    \cmidrule(lr){2-3} \cmidrule(lr){4-5}
    \multirow{5}{*}{\rotbox{Gateway\\ Offloading}}
    & \mentioned\ Security & 5  & \parbox{31em}{\mentioned\ Defining shared functionality}      & 1  \\
    & \mentioned\ Maintainability & 3 & \\
    & \mentioned\ Monitorability & 2 & \\ 
    & \notmentioned\ Performance & 1 & & \\
    & & & &\\[0.9em]
    
    \cmidrule(lr){2-3} \cmidrule(lr){4-5}
    \multirow{5}{*}[0.5em]{\rotbox{Backends for\\ Frontends}}
    & \mentioned\ Performance & 5  & \mentioned\ Complexity      & 3  \\ 
    & \mentioned\ Maintainability & 1 & & \\
    & & & &\\
    & & & &\\
    & & & &\\[0.9em]

    \cmidrule(lr){2-3} \cmidrule(lr){4-5}
    \multirow{3}{*}[0.5em]{\rotbox{Leader\\ Election}}
    & \notmentioned\ Availability & 3  & \notmentioned\ Managing leader resources      & 1  \\
    & & & &\\
    & & & &\\[0.9em]
    
    \cmidrule(lr){2-3} \cmidrule(lr){4-5}
    \multirow{5}{*}{\rotbox{Pipes and\\ Filters}}
    & \mentioned\ Performance & 5  & \notmentioned\ Monitorability      & 2  \\
    & \notmentioned\ Maintainability  & 3  & \mentioned\ Complexity & 2  \\
    & \mentioned\ Scalability & 1  &                     &    \\ 
    & \mentioned\ Reusability & 1  &                     &    \\
    & & & &\\[0.9em]

    \cmidrule(lr){2-3} \cmidrule(lr){4-5}
    \multirow{5}{*}[0.5em]{\rotbox{Static Content\\ Hosting}}
    & \mentioned\ Performance     & 5  & \notmentioned\ Eventual consistency & 1  \\
    & \mentioned\ Availability    & 3  & \mentioned\ Complexity  & 1  \\
    & \mentioned\ Maintainability & 1  &                      &    \\
    & \mentioned\ Security        & 1  &                      &    \\
    & & & &\\[0.9em]

    \cmidrule(lr){2-3} \cmidrule(lr){4-5}
    \end{tabular}
    \end{table*}
\subsubsection{\textsc{Backends for Frontends}}

Applied by five participants. All these participants note \textbf{performance} as a QA positively affected, with one participant noting that \textbf{maintainability} was also improved.
However, three participants mention costs in the form of \textbf{complexity}.

\subsubsection{\textsc{Leader Election}}

Applied by three participants. All three note \textbf{availability} as the major gain from this pattern.
As for downsides, one participant mentioned that \textbf{managing leader resources} posed a significant challenge.

\subsubsection{\textsc{Pipes and Filters}}

Applied by eight participants. \textbf{Performance} is cited by five of the participants as a gain, followed by \textbf{maintainability}, with three respondents claiming it is enhanced thanks to this pattern.
\textbf{Scalability} and \textbf{reusability} were mentioned by one participant each as additional gains.
As for drawbacks, two participants reported \textbf{difficulties monitoring} their systems with this programming model. 
Finally, \textbf{complexity} was noted by two participants as a challenge.

\subsubsection{\textsc{Static Content Hosting}}

Applied by eight participants. Over half of them declared \textbf{performance} as a benefit gained.
Following that, participants also mentioned \textbf{availability}, \textbf{maintainability} and \textbf{security} as perceived gains.
On the other hand, \textbf{eventual consistency} and \textbf{complexity} were cited as problems by one participant each.

\subsection{Measuring and addressing quality attributes} \label{sec:measuring}

Architectural trade-offs are often expressed as \textit{quality attributes} that are affected in some way by the application of a given strategy, \textit{e.g.}, the usage of a new deployment infrastructure or the application of a pattern.
Obtaining a concrete understanding of changes in software quality should ideally go beyond a developer's subjective perception, and be informed by metrics one can point to and track over time.

However, we noticed recurring participant behavior in this area: when explicitly asked for metrics for each QA, practitioners often jumped directly to techniques for addressing and improving those attributes, and we typically had to reiterate and ask them again for particular markers or indicators that can be used to measure a given attribute. 
Because of this, we were able to collect additional information that we did not originally set out to find. 
In the following sections, and through Table~\ref{tab:indicators-techniques}, for each QA we describe reported quality indicators for measuring and techniques for improving it.

\subsubsection{Scalability} \label{sec:scalability}

When it comes to getting a feel for scalability, almost half of our participants noted a close and \textbf{direct relationship to performance} (see Section~\ref{sec:performance}). This relationship is also commonly found in the literature, \textit{i.e.}, a scalable system is understood as showing increased performance in a manner proportional to the resources added. 
Two participants stated that measuring scalability is a matter of \textbf{subjective interpretation}.
One participant explicitly stated that there was no interest in measuring scalability in their company, noting that a single instance of each service was enough for their needs, even in peaks of $90\,000$ users.
Participants reported a handful of techniques for addressing scalability, most often mentioning \textbf{autoscaling}, \textit{i.e.}, using infrastructure (such as Kubernetes or AWS) that handles horizontal or vertical scaling depending on specified criteria. 
Additionally, \textbf{load testing} was reported by three participants. 
Practitioners turn to this form of execution-based testing to evaluate how their applications handle increases in requests.
\textbf{Dashboards} were mentioned by two participants as tools that visualize statistics and metrics about their applications which they turn to for guiding their business decisions (see Section~\ref{sec:monitorability}). 
\textbf{External configurations} were also noted as a form of addressing scalability, as it allows teams to autonomously reconfigure application container properties or other run-time settings.
The \textbf{segregation of responsibilities}, encouraged naturally by microservices, was raised as an important factor for addressing scalability: decoupling and isolating business concerns allows teams to scale independently and with relative ease.
Finally, \textbf{FinOps}, \textit{i.e.}, Financial Operations, was cited by one participant as a strategy to use when addressing scalability.

\begin{table*}[t]
\scriptsize
\centering
\caption{Reported indicators for measuring and techniques for addressing the seven QAs.}
\label{tab:indicators-techniques}
    \begin{tabular}[t]{
    c<{\hspace{-0.5em}}p{3.3cm}lp{3.1cm}l}
     & \textbf{Indicator} & \# & \textbf{Technique} & \# \\
    \cmidrule(lr){2-3} \cmidrule(lr){4-5}
    \parbox[t]{2mm}{\multirow{6}{*}{\rotatebox[origin=c]{90}{Scalability}}}
    & Through performance       & 4 & Autoscaling                   & 4 \\
    & Subjective interpretation & 2 & Load testing & 3 \\
    &                           & 1 & External configs              & 2  \\
    &                             &   & Dashboards & 2 \\
    &                             &   & Segregation of responsibility & 1  \\
    &                             &   &  FinOps & 1 \\[0.8em]

    \cmidrule(lr){2-3} \cmidrule(lr){4-5}
    \parbox[t]{2mm}{\multirow{4}{*}{\rotatebox[origin=c]{90}{Performance}}}
    & Infrastructure (\textit{e.g.}, \textsc{cpu}, \textsc{ram}) & 2 & Dashboards & 3 \\
    & Server (\textit{e.g.}, requests, queries)                  & 2 & Performance testing & 2 \\
    & Business performance metrics                               & 1 & &  \\
    &                                                            & 1 & &  \\[0.8em]
    
    \cmidrule(lr){2-3} \cmidrule(lr){4-5}
    \parbox[t]{2mm}{\multirow{4}{*}{\rotatebox[origin=c]{90}{Availability}}}
    & \mbox{Latency, traffic, errors, saturation} & 2 & Health checks & 3 \\
    & SLAs, SLIs, SLOs                            & 2 & Service replication & 2  \\
    &                                             & 1 & Load balancer & 1 \\
    & & & &\\[0.8em]

    \cmidrule(lr){2-3} \cmidrule(lr){4-5}
    \parbox[t]{2mm}{\multirow{5}{*}{\rotatebox[origin=c]{90}{Monitorability}}}
    & Subjective interpretation    & 1 & Third-party monitoring & 5 \\
    &                                &   & First-party monitoring & 3  \\
    & & & &\\
    & & & &\\[0.8em]

    \cmidrule(lr){2-3} \cmidrule(lr){4-5}
    \end{tabular}
\quad
    \begin{tabular}[t]{c<{\hspace{-0.5em}}p{3.2cm}lp{3.1cm}l}
     & \textbf{Indicator} & \# & \textbf{Technique} & \# \\
     
    \cmidrule(lr){2-3} \cmidrule(lr){4-5}
    \parbox[t]{2mm}{\multirow{6}{*}{\rotatebox[origin=c]{90}{Security}}}
    & Code security & 1 & Internal audits & 5 \\
    &                 &   & Penetration testing & 3 \\
    &                 &   & External audits & 2 \\
    &                 &   & Bug bounties & 1 \\
    &                 &   & Multi-tenancy & 1 \\
    &                 &   & SIEM        & 1 \\[0.8em]

    \cmidrule(lr){2-3} \cmidrule(lr){4-5}
    \parbox[t]{2mm}{\multirow{8}{*}{\rotatebox[origin=c]{90}{Testability}}}
    & Number of automated tests & 3 & Unit testing & 8 \\
    & Code coverage             & 3 & Integration testing & 4 \\
    &                           &   & Performance testing & 3  \\ 
    &                           &   & Functional testing & 2 \\
    &                           &   & Penetration testing & 1 \\
    &                           &   & Smoke testing & 1 \\
    &                           &   & Non-regression testing & 1 \\ 
    &                           &   & QA Manifesto & 1 \\[0.8em]

    \cmidrule(lr){2-3} \cmidrule(lr){4-5}
    \parbox[t]{2mm}{\multirow{5}{*}{\rotatebox[origin=c]{90}{Maintainability}}}
    & Code coverage      & 2 & Static code analysis & 4 \\
    & Time to fix bugs   & 1 & Automated pipelines & 3 \\
    & Architectural debt & 1 & Code review & 3 \\
    &                      &   & Domain-driven design & 2 \\
    & & & &\\[0.5em]
    
    \cmidrule(lr){2-3} \cmidrule(lr){4-5}

    \end{tabular}
\end{table*}

\subsubsection{Performance} \label{sec:performance}

With performance, participants were aware of lower-level metrics for assessing the current level, namely \textbf{hardware or infrastructure properties} (\textit{e.g.}, CPU and RAM usage) and \textbf{server properties} (\textit{e.g.}, number of active requests or time taken handling requests, or database queries).
Besides technical metrics, one participant referred to \textbf{business performance metrics}, also known as key performance indicators, as an additional instrument for evaluating performance.
One participant stated explicitly that his company has no interest in run-time performance, as it poses little cause for concern for their particular business.
For addressing performance, participants mention \textbf{dashboards} the most, with 3 mentioning these tools as the go-to for assessing performance at a glance---for more details on dashboards, see Section~\ref{sec:monitorability}. 
Dashboards effectively synthesize a range of indicators into one, sitting at a higher level of abstraction.
Furthermore, \textbf{performance testing} was reported by two participants as a strategy for measuring performance, integrated into continuous deployment pipelines.

\subsubsection{Availability}

For measuring availability, two participants noted run-time statistics like \textbf{latency, traffic, errors and saturation} as ways to evaluate availability.
Additionally, two participants mention tracking Service Level Agreements, Service Level Objectives, and Service Level Indicators for judging availability (\textbf{SLAs, SLOs, and SLIs}, respectively). 
One participant reported explicitly no interest in availability due to not having concerns for critical or otherwise real-time processes in their business.
As for tackling availability, performing \textbf{health checks} received most mentions from three respondents, who all noted having automatic alerts set up to notify them when their systems are down. 
Furthermore, \textbf{service replication} and the usage of a \textbf{load balancer} were found, as expected.

\subsubsection{Monitorability} \label{sec:monitorability}

While most participants ignored the question of \textit{how they measure monitorability}---jumping straight to the tools they use which aim to address it---one participant told us that, while an essential aspect, getting a feel for monitorability is a matter of professional intuition and experience.
As for addressing monitorability, most respondents cited using \textbf{third-party monitoring} for assisting their observing needs.
Some tools were mentioned several times.
Grafana and the ELK stack (Elasticsearch, Logstash and Kibana) were the more popular ones (each with 4 mentions), followed by Prometheus (3 mentions) and Google Analytics (2 mentions).

Other third-party tools were mentioned by only one participant each: VictorOps, Opsgenie, Datadog, Dynatrace, Microsoft Power BI, and Google Data Studio.
In addition, three respondents explicitly mentioned developing \textbf{first-party monitoring} solutions. %
One main driver cited for developing an in-house monitoring infrastructure is to avoid vendor lock-in.

\subsubsection{Security}

Security seemed a puzzling question for our participants, with only one pointing out any concrete metrics, in the form of \textbf{code security}.
As a key part of a business, practitioners are, however, very interested in addressing security.
They do this via \textbf{internal audits}, which received most mentions from participants, followed by \textbf{penetration testing}, noted by three participants.
Our two participants in the finance industry additionally mention \textbf{external audits} from entities that issue PCI certificates, required for these companies to stay in business.
Finally, other techniques like \textbf{security information and event management (SIEM)}, \textbf{bug bounties} and \textbf{multi-tenancy} were mentioned by one participant each as approaches to ensure better security in their business.

\subsubsection{Testability}

For measuring testability, having a \textbf{number of automated tests} within DevOps pipelines and \textbf{code coverage} are the go-to indicators for practitioners to judge this QA.
When describing how they addressing it, \textbf{unit testing} receives most mentions from participants, followed by \textbf{integration testing}, \textbf{performance testing}, \textbf{penetration testing} and \textbf{functional testing}. 
One participant specified, in addition, the usage of \textbf{smoke testing} and \textbf{non-regression testing}. 
The same interviewee commented that he felt it was the most neglected QA out of the set.
One participant mentioned their company having a \textbf{QA Manifesto}, an internal document containing their definition of quality revolving around tests.

\subsubsection{Maintainability}

Regarding metrics for maintainability, participants were more divided. 
Similar to testability, \textbf{code coverage} was mentioned by two participants as an indicator of maintainability. 
Beyond that, one participant mentioned tracking, through Slack and Jira, bug reports and the \textbf{time taken to resolve issues} as a recent effort started in the company aimed at upholding a desired level of maintainability.
Finally, one other participant reported an analysis of \textbf{architectural debt} as a measure of levels of maintainability.
As for increasing maintainability, \textbf{static code analysis} seems to be the preferred strategy, with three participants mentioning using SonarQube and one using Codacy for this effect.
Following that, \textbf{automated pipelines} and \textbf{code reviews} were also cited three times as fundamental aspects of higher maintainability.
Finally, \textbf{domain-driven design}, which affirms a strict logical match between business concerns and the structure of software systems, was mentioned by two participants as one of the ways they can ensure a better level of maintainability.

\section{Discussion} \label{sec:discussion}

The next sections present our interpretation and discussion of the results described in Section~\ref{sec:findings} and identify the main threats to validity and strategies to their mitigation. 

\subsection{Pattern trade-offs}

As mentioned in Section~\ref{sec:tradeoff}, our findings directly mix QAs and more concrete concerns such as, for example, ``\textit{Managing data consistency}". 
We refrained from approximating these instances to QAs which \textit{best matched}---\textit{e.g.}, using ``\textit{Maintainability}" in place of ``\textit{Complexity}"---for one main reason: such a mapping would invariably produce ambiguity and biases which could negatively impact our (and the reader's) interpretation of the results.

All in all, we were able to gather many insights into each of these particular patterns, some of which differ quite substantially from their original intention documented in the AAC.
Some concerns were raised in the AAC but not by our participants. 
We did not raise these concerns during the interview, so the professional perception of their severity is left unclear. 
However, their lack of mention suggests they are minor or secondary issues.

In several cases, we unfortunately could not discuss some concepts in-depth due to a mix of time constraints, differences in how the unstructured segments of the interviews unfolded, and participants not having had experience with some of the patterns, impairing our ability to extract more findings. 

In any case, \textbf{we identified 14 new gains and 6 new pains} for these patterns that were not present in the AAC documentation, out of a total of 35 gains and 25 pains reported by participants. 
This suggests that the knowledge in the AAC may be incomplete. 
\textsc{Compute Resource Consolidation} was the only pattern without any pains reported. 

In a scenario where all 14 patterns are applied, we gain benefits in the form of:
\textit{maintainability} from 7 patterns;
\textit{performance} from 6 patterns;
\textit{scalability} from 6 patterns;
\textit{availability} from 5 patterns;
\textit{security} from 5 patterns;
\textit{monitorability} from 3 patterns; 
and other gains from 1 pattern each.
On the other hand, we are also afflicted with:
\textit{complexity} from 5 patterns;
\textit{single point of failure} from 3 patterns,
\textit{operational complexity} from 2 patterns;
and other pains from 1 pattern each.

\subsection{Measuring and addressing quality attributes}

The results reported in Section~\ref{sec:measuring}, concerning how QAs are measured and addressed, were not possible to obtain for all participants. We are unclear as to the reasons for this, but we do not think it is due to these practitioners not actively measuring a given QA.
In those cases, the interviewees briefly described, instead, the importance they gave to that attribute, from their personal or company's perspective. 
We omit these remarks from our findings due to their general irrelevance to our RQs and lack of depth.
These hurdles give us less confidence in asserting saturation of our findings, though we do feel we provide a relevant account of industrial perception of software QAs.

Furthermore, it is possible that metrics for measuring certain QAs went unreported simply due to being obvious to the point of second nature. 
For example, response time is an immediately obvious metric for assessing scalability or performance (to an extent) yet it was not mentioned explicitly by any participant.
This is likely due to already being accounted for by measuring tools like dashboards, and, as such, does not warrant special mention from respondents.

Overall, we found that participants do not measure software qualities directly. 
They correctly conceive of each individual quality as a sum of lower level attributes, which they can then project onto a whole, when needed. 
But, due to the nature of commercial software development, software engineers want measures for QAs that are easy to check, and they want simple procedures to confirm that the readings are consistent.
As a result, we were left with the impression that QAs are not part of the daily vocabulary of the software developer, likely due to an insipid and highly personal conceptual understanding of each attribute, hampering interpersonal agreements on a semantic basis. 
All in all, these findings seem to support the arguments presented by Moses~\cite{moses2009should}, particularly the author's note of skepticism regarding industry practice: ``\textit{software developers and researchers may remain content to measure attributes or processes considered to be related to quality attributes}".

Furthermore, when we compare our findings with the work by Li~\cite{li_understanding_2021}, we see several techniques practitioners resort to for addressing these QAs that were not covered by their systematic literature review. 

\subsection{Threats to validity} \label{sec:threats}

The credibility of results obtained through empirical studies is always hampered by threats to validity~\cite{Wedyan_Abufakher_2020}.
We can think of validity through four different aspects~\cite{runeson2009guidelines}: construct validity; internal validity; external validity; and reliability. 

\textbf{Construct validity.}  An interview study is susceptible to social desirability bias, which refers to the tendency for participants to present themselves in the most flattering light~\cite{Wedyan_Abufakher_2020}.
To mitigate this threat, we assured participants that no identifying data would be shared.

\textbf{Internal validity.} Through our previous interest in the field, we were pre-exposed to existing research. This might have biased our interview design and may have led to some elements going unnoticed. 
For instance, we targeted our interview questions towards the impact patterns have on explicitly mentioned or obviously implicit QAs, which may have left some attributes unaddressed.
Further, this study demanded extensive knowledge of all design patterns and QAs that were analyzed, as well as a proficiency for conveying this knowledge to participants to prevent miscommunication or confusion. 
While we sought a serious depth of knowledge of the theoretical elements involved in this study, all interviews were performed by one author, who cannot claim extensive expertise of the domain of software architecture. 
As a result, it is possible that degradation of communication occurred with some respondents or misinterpretation on our part, negatively affecting the efficiency of the interviews.
Finally, we cannot claim completeness of either the set of identified architectural trade-offs of design patterns or the set of indicators for measuring and techniques for addressing QAs.

\textbf{External validity.} We cannot claim saturation of our findings for the patterns with a low adoption rate. Especially for these patterns, nine interviews is clearly not enough to draw conclusions that can be confidently generalized to the software industry in general.
Furthermore, voluntary surveys are inherently susceptible to self-selection bias~\cite{ayas2021facing} as uninterested respondents are unlikely to participate.
Finally, all interviewed participants are currently residing and working in Portugal. As such, it is theoretically possible that a regional technology culture might have rendered a biased lens into the industrial reality of the microservices architecture.
To mitigate this threat, we selected participants from diverse backgrounds to cover companies of different sizes and business domains.

\textbf{Reliability.} Many of steps described in this study were performed by only one of the researchers. It is possible that different people might have drawn different answers during the interviews, or might have made different options during coding. We may still improve the latter in subsequent steps of our work, by having coding done by a second researcher, and bring in a third researcher to solve disagreements.

\section{Related Work} \label{sec:related}

Our work combines two relevant aspects in microservices design: the use of patterns and software quality. While both topics have already been addressed in scientific literature, studies examining their interplay are still rare. We therefore discuss the studies that make a connection between both aspects and highlight important empirical investigations that reflect the current industry practices in this regard.

Osses \textit{et al.}~\cite{osses_exploratory_2019} extracted 44 patterns and tactics used for microservices from 1,067 studies in a systematic review. In a related study~\cite{Osses2018a}, the authors focus on the intersection of patterns and tactics that have been described in academia as well as in industry. The authors found that most architectural patterns are associated with five quality attributes: scalability, flexibility, testability, performance, and elasticity. As another result, they propose a pattern taxonomy for microservices. Together with the review by Li \textit{et al.}~\cite{li_understanding_2021}, who sought a comprehensive understanding of QAs in the context of the microservices architecture, these contributions build a solid foundation for our work. The authors of the latter study also affirm a need for further empirical research addressing particular QAs like \textit{maintainability}.

Valdivia \textit{et al.}~\cite{valdivia_quality_2019} note the unclear mapping between QAs and patterns in microservices and sought to address this issue by a systematic literature review. While the authors present a comprehensive compilation of patterns with their related QAs, the impact of the QAs is not investigated. In this regard, the authors state that selecting certain patterns requires an analysis of the trade-offs specifically for the system in question, as their impact could not be generalized.
This inherent difficulty is confirmed by Rosa \textit{et al.}~\cite{rosa_method_2020}, who identified a lack of knowledge about patterns among software architects and, as mitigation, developed a method for an architectural trade-off analysis based on patterns in microservices. The authors, however, only investigate the relationship between patterns and structural attributes, and do not target QAs in particular.

Besides these works, several empirical studies have been conducted, including a controlled experiment on the impact of service-oriented patterns on software evolvability~\cite{bogner_impact_2019} and a survey on the adoption of patterns for engineering cloud software~\cite{Sousa_Ferreira_Correia_2021}. Taibi \textit{et al.}~\cite{Taibi2019b} surveyed 23 publications on industry case studies and thereby identified widely adopted architectural principles and patterns for microservices, including their trade-offs.

The presented studies cover the analysis of architectural properties and design trade-offs for microservices reasonably well. An adherence to patterns is commonly desired in professional software development and can be an effective facilitator for creating systems with a set of desired software qualities. However, empirical evidence is needed to confirm this hypothesis. The discussed studies have rarely attempted to evaluate the consequences that patterns have on particular QAs, and the few that did have generally resorted to reviewing academic literature or analyzed a small set of patterns and QAs.
As the existing studies did not sufficiently address this topic, this gap in scientific research sets the motivation for our study. The work by Valdivia \textit{et al.}~\cite{valdivia_quality_2019} took the first steps in this direction but solely relied on existing literature. To the best of our knowledge, there exists no empirical study reflecting the perception by industry on design patterns in the microservices architecture and their quality impacts---positive and negative.

\section{Conclusion} \label{sec:conclusion}

We began with the observation that the microservices architecture is an increasingly accepted and adopted architectural style, as it helps overcome the limitations of traditional monolithic systems, namely when it concerns issues of scalability, crucial when moving to a cloud-first paradigm. 
Designing these systems, however, is no easy task, and the careful consideration of which design patterns to apply in the context of large-scale distributed systems has been subject to little rigorous empirical research. Our RQs, initially posited in Section~\ref{sec:approach}, can now be answered the following way:

\begin{itemize}[leftmargin=*]
    \item \textbf{RQ1:} \textbf{\textit{What is the rationale for the adoption of patterns in microservices systems?}} 
    --- 
    Practitioners are aware of the problems presented, but they do not think of their solutions in terms of patterns. 
    Oftentimes, participants would state they are familiar with the solution described in a pattern but unaware that there was such a name for that solution.
    This suggests that some professionals may currently lack the language to explore solutions for hurdles seen in microservices, which might lead to risky and untested paths being taken unnecessarily.
    Queried for the expertise regarding design patterns, each practitioner had experience with, on average, 9 of the 14 patterns we showed them, with only one participant having had experience with all. 
    \textsc{Strangler} and \textsc{Gateway Routing} were the only patterns adopted by all respondents.
    
    \item \textbf{RQ2:} \textbf{\textit{How are QAs influenced as a result of applying microservice patterns?}} 
    --- 
    Practitioners' perceptions of the architectural trade-offs inherent to the design patterns we asked about largely matched the original documentation, implying that these patterns are a reliable resource.
    However, new gains and pains were identified in some patterns as a result of this study, suggesting that they may be incomplete or not completely accurate. 
    For example, \textsc{Gateway Aggregation} was viewed very poorly by all participants who applied it, leading us to hypothesize that the gains this pattern may provide are not worth the costs.
    On the other hand, \textsc{Compute Resource Consolidation} was viewed very positively, with no participants reporting any considerable pains related to it.
    
    \item \textbf{RQ3:} \textbf{\textit{How are QAs measured in microservices?}} 
    --- 
    Practitioners have shown interest in tracking key facets of software quality, though they do not often turn to QAs as the terminology to express and measure characteristics of their systems. Only a few well-known terms emerged in our results, like \textit{scalability}, \textit{performance}, and \textit{maintainability}.
    Similarly, practitioners have shown little interest in discussing concrete metrics. 
    Instead, they directly think in terms of techniques and tools. 
\end{itemize}

The results and findings from this article offer implications for both professionals and researchers.
For professionals, this work provides further understanding of QAs impacted by patterns, which can serve as a useful guide for strategic engineering decision-making on moving forward with adopting microservices.
For researchers, we provide a report of an interview study's design and execution, with findings that  support new hypothesis of how the existing body of knowledge may be incomplete or inaccurate.

\section{Future work}

Further empirical research may provide more insights regarding the methods for designing microservice systems. Namely, applying this study's methodology using a larger sample size will yield more reliable results, and studies specifically targeting quality attributes in microservices are also sorely needed.
Additionally, a systematic review of the empirical literature on microservices that aggregates knowledge on pains and gains would further assist researchers in obtaining an actualized perspective on the design trade-offs inherent to this architectural style. In this article, we restrict our analysis of related works (\textit{cf}. Section~\ref{sec:related}) to those addressing simultaneously patterns and software quality aspects in the scope of microservices design, but such a systematic review could also look in more detail into works as those done by Li \textit{et al.}~\cite{li2022enjoy}, Reis \textit{et al.}~\cite{reis2021developing}, Bogner \textit{et al.}~\cite{bogner2019assuring,bogner2019microservices}, Soldani \textit{et al.}~\cite{soldani2018pains} and Waseem \textit{et al.}~\cite{waseem2021design}, which study practices and quality aspects in the context of microservices. Even if not necessarily leaning on pattern literature and on how patterns are perceived by professionals, we believe that some of the results of these studies could be used to improve existing pattern knowledge.

\section*{Acknowledgement}

The authors would like to thank Nour Ali for her insightful comments on an early version of this work. 

\newpage

% References
\bibliographystyle{IEEEtran}
\bibliography{text}

\end{document}